\documentclass[aps,prl,twocolumn,nofootinbib,a4paper,amsfonts,amssymb,amsmath,floatfix,superscriptaddress]{revtex4-2}

\pdfoutput=1
\usepackage{graphicx}
\usepackage[section]{placeins}
\usepackage{amsmath}
\usepackage{amsthm}
\usepackage{amssymb}
\usepackage{dsfont}
\usepackage{bbold}
\usepackage{color}
\usepackage{comment}
\usepackage{appendix}
\usepackage{dutchcal}
\usepackage{tikz}
\usepackage{subcaption}
\usepackage{url}

\newtheorem{theorem}{Theorem}
\newtheorem{lemma}{Lemma}

\newcommand{\Tr}{\text{Tr}}

\newcommand\id{\leavevmode\hbox{\small1\kern-3.3pt\normalsize1}}
\newcommand{\ack}{\subsection*{\normalsize \sf \textbf{Acknowledgment}}}
\usepackage{braket}

\usepackage[colorlinks=true,allcolors=magenta,breaklinks=true]{hyperref}

\begin{document}

\title{
Generalizing Bell nonlocality without global causal assumptions
}
	
\author{Ravi Kunjwal}
\email{quaintum.research@gmail.com}
\affiliation{Universit\'e libre de Bruxelles, QuIC, Brussels, Belgium}
\affiliation{Aix-Marseille University, CNRS, LIS, Marseille, France}

	\author{Ognyan Oreshkov}
	\email{ognyan.oreshkov@ulb.be}
\affiliation{Universit\'e libre de Bruxelles, QuIC, Brussels, Belgium}
	
	\date{\today}                                           
	
	\begin{abstract}
	
Bell scenarios are multipartite scenarios that exclude any communication between parties.
This constraint leads to a strict hierarchy of correlation sets in such scenarios, namely, classical, quantum, and nonsignaling. However, without any constraints on communication  between the parties, they can realize arbitrary correlations by exchanging only classical systems. Here we consider a multipartite scenario where the parties can engage in at most a \textit{single} round of communication, \textit{i.e.}, each party is allowed to receive a system once, implement any local intervention on it, and  send out the resulting system once. While no global assumption about causal relations \textit{between} parties is assumed in this scenario, we do make a causal assumption \textit{local} to each party, \textit{i.e.}, the input received by it causally precedes the output it sends out.  
We then introduce  \textit{antinomicity}, a notion of nonclassicality for correlations in such scenarios, and prove the existence of 
a strict hierarchy of correlation sets classified by their antinomicity. Antinomicity serves as a generalization of Bell nonlocality: when all the parties discard their output systems (\textit{i.e.}, in a nonsignaling scenario), it is mathematically equivalent to Bell nonlocality. Like Bell nonlocality, it can be understood as an instance of fine-tuning, one that is necessary in any classical model of cyclic causation that avoids time-travel antinomies but allows antinomic correlations. Furthermore, antinomicity resolves a long-standing puzzle, \textit{i.e.}, the failure of causal inequality violations as device-independent witnesses of nonclassicality. Antinomicity implies causal inequality violations, but not conversely.
	\end{abstract}
\maketitle
The scientific enterprise hinges on uncovering causal explanations for observed correlations. Usually these explanations assume definiteness of causal order between the relevant events. If, however, causal order is subject to quantum indefiniteness in a similar sense as physical properties like position and momentum, what would such explanations look like? Aside from its intrinsically foundational motivations, this question is also motivated by considerations of what a theory of quantum gravity that combines the probabilistic aspects of quantum theory with the dynamical aspects of general relativity might look like \cite{Hardy05,Hardy07,Hardy16}. The process-matrix framework \cite{OCB12}, in offering a possible answer \cite{BLO21} to this question, allows for the violation of statistical constraints called causal inequalities that hold under the assumption of definite causal order. However, the framework has a puzzling feature: causal inequality violations occur even in its classical deterministic limit \cite{BW16}, contrary to expectations that they witness nonclassicality akin to Bell inequality violations.

To address this puzzle, we consider arbitrary multipartite correlations in a single-round communication scenario. That is, unlike Bell scenarios, we allow parties to communicate via exchanging systems in a single round, as envisioned at the operational level in the process-matrix framework \cite{OCB12} but without assuming a specific theory that gives rise to the correlations. By a `single round', we mean that the maximum number of times any party can receive or send a system, under the constraint that receiving precedes sending according to a definite causal order local to the party, is one. We then ask whether there is a natural generalization of Bell inequalities---in the sense of witnessing nonclassicality---in such scenarios without making global causal assumptions. To answer this question we introduce \textit{antinomicity} \cite{KO23}, a notion of nonclassicality inspired by Bell nonlocality but applicable without any assumption of definite causal order between different parties. If one restricts attention to scenarios where the parties discard their output systems after making their local interventions (\textit{i.e.}, they do not communicate), antinomicity formally reduces to Bell nonlocality.\footnote{We note that a different route to generalizing Bell nonlocality via partial causal constraints was recently developed by Gogioso and Pinzani \cite{GP23_geom}. Since the axiomatic foundations of their framework are very different from ours, it remains unclear whether antinomicity---which does not rely on any global causal constraints---admits an interpretation within their framework.}

Intuitively, antinomicity captures the fact that some correlations can be so strong that any classical causal explanation of them---even one invoking cyclic causality \cite{BLO21}---would necessarily hit a roadblock: Namely, the underlying classical physics must entertain time-travel antinomies \cite{BT21} that are hidden at the operational level via a statistical fine-tuning \cite{WS15}. This is analogous to how some nonsignaling correlations can be so strong that any classical causal explanation of them must---under the assumption of free local interventions\footnote{Thereby excluding the possibility of measurement settings being fixed by the local hidden variable, \textit{e.g.}, as in  superdeterminism \cite{Bell76, SEP_Bell24}.}---either allow a party to causally influence its own past (`retrocausality') \cite{Price94} or it must invoke hidden superluminal causal influences (`nonlocality') \cite{WS15}. Such influences, nevertheless, are not observed  operationally because of fine-tuning \cite{WS15}.

We  classify correlations in any single-round communication scenario into four distinct sets based on their antinomicity and prove strict inclusions between them. This is analogous to the strict inclusions between classical, quantum, and nonsignaling correlations in Bell scenarios. We identify antinomic correlations that \textit{cannot} be realized with process matrices, analogous to the unachievability of PR boxes \cite{PR94} with nonsignaling quantum correlations. More specifically, all deterministic correlations that are not realizable via process functions \cite{BW16,BW16fp,BT21} (and which are therefore antinomic, see Theorem \ref{thm:det} below) cannot be realized by process matrices. One of these, namely, the correlation that wins the bipartite \textit{`Guess Your Neighbor's Input'} (GYNI) game \cite{BAF15} perfectly, was previously shown to be unachievable in any finite dimension via a dimension-dependent bound \cite{BAB19}. Moreover, following results we report here \cite{KO23}, Liu and Chiribella also obtained Tsirelson-type bounds on some causal inequalities \cite{LC25}.

While in bipartite scenarios antinomicity is equivalent to causal inequality violations \cite{OCB12}, this equivalence doesn't hold for three or more parties, \textit{i.e.}, antinomicity is sufficient for causal inequality violations but not necessary. To clarify this distinction, we propose a tripartite antinomicity witness, termed the \textit{`Guess Your Neighbor’s Input, or NOT'} (GYNIN) inequality. We show that this inequality cannot be saturated by causal correlations and is maximally violated by a process matrix.

For approaches to quantum gravity \cite{Hardy05,OCB12} that allow for causal inequality violations without time-travel antinomies \cite{BT21,Godel49,Luminet21}, our results could provide an operational way to determine whether such causal indefiniteness requires a nonclassical spacetime. That is, experimenters in local labs can, based only on the antinomicity of observed correlations, rule out the possibility that their labs are embedded in an environment modelled by classical closed timelike curves without time-travel antinomies (known as process functions \cite{Baumeler17, BT21}). Since we make minimal assumptions, antinomicity can provide a tool to discriminate between different models of nonclassical spacetimes \cite{GB22} in terms of their ability to support antinomic correlations. This is analogous to how Bell nonlocality---namely, observed correlations in a Bell scenario that cannot arise from classical shared randomness---provides a useful tool to classify nonsignaling theories according to their ability to support Bell nonlocal correlations \cite{MAG06,Barrett07,BCP14}.

We now define some preliminary concepts before moving on to our main results.

\textit{The operational paradigm.---}We work within the following operational paradigm (introduced in \cite{OCB12}): consider $N$ isolated labs embedded in some environment; each lab can receive an input system from the environment and subsequently (according to a local notion of definite causal order) send an output system to the environment; party $S_k$ (in the $k$th lab) receives a classical setting (or question), $a_k\in A_k$, and reports a classical outcome (or answer), $x_k\in X_k$. To determine $x_k$, $S_k$ can implement some local intervention based on $a_k$ on the input system it receives from the environment and, depending on the result of this local intervention, $S_k$ reports the answer $x_k$ and sends an output system to the environment. Crucially, each party can apply arbitrary local interventions on the input system it receives. The communication between the labs is mediated entirely by the environment, with the parties limited to local interventions in their respective labs. The central object of investigation is the observed multipartite correlation $p(\vec{x}|
\vec{a}):=p(x_1,\dots,x_N|a_1,\dots,a_N)$. Note that the parties can execute at most a \textit{single-round} communication protocol since each party receives or sends out a system at most once (see Ref.~\cite{HO21} for a generalization to multiple rounds).\footnote{Our operational paradigm specializes to a Bell scenario if the parties discard their output systems, \textit{i.e.}, they do not communicate and merely receive input systems once from the environment.} This operational paradigm is  \textit{a priori} agnostic about i) the nature of the input/output systems received/sent by the local labs, ii) the nature of the local interventions on them, and iii) the nature of communication between the labs that is mediated by the environment. To specify the nature of these three aspects amounts to specifying a particular operational theory that fixes them, \textit{e.g.}, the (quantum) process-matrix framework, where the input/output systems are quantum systems, the local interventions are quantum instruments, and the nature of (quantum) communication between the labs is dictated by the process describing the environment. In general, the  specification of a particular operational theory restricts the set of multipartite correlations $p(\vec{x}|\vec{a})$ to a subset of the set of all correlations.

\textit{The process-matrix framework.---}Within the operational paradigm outlined above, we now assume that the parties perform local \textit{quantum} operations, \textit{i.e.}, party $S_k$ has an incoming quantum system $I_k$ with Hilbert space $\mathcal{H}^{I_k}$, an outgoing quantum system $O_k$ with Hilbert space $\mathcal{H}^{O_k}$, and can perform arbitrary quantum operations from $I_k$ to $O_k$. A local quantum operation is described by a quantum instrument, \textit{i.e.}, a set of completely positive (CP) maps $\{\mathcal{M}^{S_k}_{x_k|a_k}:\mathcal{L}(\mathcal{H}^{I_k})\rightarrow\mathcal{L}(\mathcal{H}^{O_k})\}_{x_k\in X_k}$, where the setting $a_k\in A_k$ labels the instrument and $x_k\in X_k$ labels the classical outcome associated with each CP map.\footnote{Without loss of generality, we assume that the outcome set $X_k$ is identical for all settings $a_k\in A_k$: this can be ensured by including, if needed, outcomes that never occur, \textit{i.e.}, those represented by the null CP map, for settings that have fewer non-null outcomes than some other setting.} Summing over the classical outcomes yields a completely positive and trace-preserving (CPTP) map $\mathcal{M}^{S_k}_{a_k}:=\sum_{x_k}\mathcal{M}^{S_k}_{x_k|a_k}$. The correlations between the classical outcomes of the different labs given their classical settings are given by
\begin{align}\label{eq:corrpm}
	&p(x_1,x_2,\dots,x_N|a_1,a_2,\dots,a_N)\nonumber\\
	=&\Tr(W M^{I_1O_1}_{x_1|a_1}\otimes M^{I_2O_2}_{x_2|a_2}\otimes\dots\otimes M^{I_NO_N}_{x_N|a_N}),
\end{align}
where 
	$M_{x_k|a_k}^{I_kO_k}:=[\mathcal{I}^{I_k}\otimes\mathcal{M}^{S_k}_{x_k|a_k}(d_{I_k}\ket{\Phi^+}\bra{\Phi^+})]^{\rm T} \in\mathcal{L}(\mathcal{H}^{I_k}\otimes\mathcal{H}^{O_k})$
is the Choi-Jamiołkowski (CJ) matrix associated to the CP map $\mathcal{M}^{S_k}_{x_k|a_k}$, $\ket{\Phi^+}\in\mathcal{H}^{I_k}\otimes\mathcal{H}^{I_k}$ is the maximally entangled state, \textit{i.e.}, $\ket{\Phi^+}=\frac{1}{\sqrt{d_{I_k}}}\sum_{j=1}^{d_{I_k}}\ket{j}\ket{j}$, and $\mathcal{I}^{I_k}:\mathcal{L}(\mathcal{H}^{I_k})\rightarrow\mathcal{L}(\mathcal{H}^{I_k})$ is the identity channel. We have that
	$M_{a_k}^{I_kO_k}\geq 0$, $\Tr_{O_k}M_{a_k}^{I_kO_k}=\id_{I_k}$.

The operator $W\in\mathcal{L}(\mathcal{H}^{I_1}\otimes\mathcal{H}^{O_1}\otimes\mathcal{H}^{I_2}\otimes\mathcal{H}^{O_2}\otimes\dots\otimes\mathcal{H}^{I_N}\otimes\mathcal{H}^{O_N})$ is called a process matrix and it establishes correlations between the local interventions of the labs. $W$ satisfies the following constraints: $W\geq 0$, and $\forall M^{S_k}\geq 0$ where $M^{S_k}\in\mathcal{L}(\mathcal{H}^{I_k}\otimes\mathcal{H}^{O_k})$ and $\Tr_{O_k}M^{S_k}=\id_{I_k}: \Tr(W \bigotimes_{k=1}^N M^{S_k})=1$.

\textit{Classical processes and process functions.}---We now consider two other instantiations of the operational paradigm, the \textit{classical process framework} and the \textit{process-function framework}, both of which can be shown to arise in the diagonal limit of the process-matrix framework (\textit{i.e.}, where the process matrix is assumed to be diagonal relative to some fixed choice of bases) \cite{BW16}. 
In both cases, the input and output systems are classical random variables, the local operation of each party is a classical stochastic map, and the environment is described by a conditional probability distribution dictating how the inputs received by the parties are affected by the outputs sent out by the parties.

More concretely, each party $S_k$ has an incoming classical system represented by a random variable $I_k$ that takes values $i_k\in\{0,1,\dots,d_{I_k}-1\}$ and an outgoing classical system represented by a random variable $O_k$ that takes values $o_k\in\{0,1,\dots,d_{O_k}-1\}$.\footnote{With a slight but standard abuse of notation, we will often also use $I_k$ and $O_k$ to represent the respective sets in which these random variables take values.} The local operations of party $S_k$ are specified by the conditional probability distribution $p(x_k,o_k|a_k,i_k)\in[0,1]$, where $a_k$ and $x_k$ denote, respectively, the setting and outcome for party $S_k$. Using the notation $\vec{x}:=(x_1,x_2,\dots,x_N)$, and  $\vec{a}:=(a_1,a_2,\dots,a_N)$, the multipartite correlations, $p(\vec{x}|\vec{a})$, are then given by 
\begin{align}\label{eq:loccons1}
	p(\vec{x}|\vec{a})=\sum_{\vec{i},\vec{o}}\prod_{k=1}^Np(x_k,o_k|a_k,i_k)p(\vec{i}|\vec{o}),
\end{align}
where $p(\vec{i}|\vec{o})$ is the conditional probability distribution describing the environment and, as such, is not arbitrary but constrained to satisfy the requirement of \textit{logical consistency}, \textit{i.e.}, $p(\vec{i}|\vec{o})$ should be such that for any arbitrary choices of local interventions by the parties, $\{p(x_k,o_k|a_k,i_k)\}_{k=1}^N$, the correlation defined by Eq.~\eqref{eq:loccons1} satisfies non-negativity ($p(\vec{x}|\vec{a})\geq 0$ for all $\vec{x},\vec{a}$) and  normalization ($\sum_{\vec{x}}p(\vec{x}|\vec{a})=1$ for all $\vec{a}$). This condition of logical consistency is necessary and sufficient to exclude the possibility of time-travel antinomies \cite{BW16,BT21}. A logically consistent $p(\vec{i}|\vec{o})$ is a \textit{classical process} in the sense of Ref.~\cite{BW16} and correlations that are achievable by such a process via Eq.~\eqref{eq:loccons1} are said to belong to the \textit{classical process framework}.

On the other hand, correlations achievable via classical processes of the following form are said to belong to the \textit{process-function framework} \cite{BW16,BT21}: $p(\vec{i}|\vec{o})=\sum_{\lambda}p(\lambda)\delta_{\vec{i},\omega^{\lambda}(\vec{o})}$, where $\lambda$ labels the process function $\omega^{\lambda}:\vec{O}\rightarrow\vec{I}$ \cite{BW16, BW16fp, BT21, TB23} defined via $\omega^{\lambda}:=(\omega^{\lambda}_k:\vec{O}\rightarrow I_k)_{k=1}^N$. A process function is a map from the outputs of the parties to their inputs that satisfies logical consistency when written as a conditional probability distribution $p_{\lambda}(\vec{i}|\vec{o}):=\delta_{\vec{i},\omega^{\lambda}(\vec{o})}$. The set of classical processes defining the process-function framework corresponds to the \textit{deterministic-extrema polytope} of Ref.~\cite{BW16}.

\textit{Classical quasi-processes.---}We will refer to an arbitrary conditional probability distribution $p(\vec{i}|\vec{o})$ as a \textit{classical quasi-process} and when this distribution is deterministic, we will represent it via a \textit{quasi-process function} $\omega:\vec{O}\rightarrow\vec{I}$, where  $\omega:=(\omega_1,\omega_2,\dots,\omega_N)$, $\omega_k:\vec{O}\rightarrow I_k$ for all $k\in\{1,2,\dots,N\}$, and $p(\vec{i}|\vec{o})=\delta_{\vec{i},\omega(\vec{o})}=\prod_{k=1}^N\delta_{i_k,\omega_k(\vec{o})}$.\footnote{This is non-standard terminology, but we will later find it useful in describing the most general set of correlations in multipartite scenarios.} A classical quasi-process that satisfies logical consistency is a \textit{classical process} \cite{BW16}. If a classical process is deterministic, the quasi-process function associated with it is a \textit{process function} \cite{BT21}. The correlations achievable by a classical quasi-process $p(\vec{i}|\vec{o})$ are given by Eq.~\eqref{eq:loccons1}, with the caveat that local interventions cannot be arbitrary when the classical quasi-process fails to be a classical process, \textit{i.e.}, some local interventions must be disallowed in that case to ensure that the left-hand-side of Eq.~\eqref{eq:loccons1} is a conditional probability distribution that is normalized for all settings. This restriction on local interventions means that classical quasi-processes do not, in general, fall within the operational paradigm we envisage.\footnote{A concrete example outside our operational paradigm is the single-party quasi-process function $p(i_1|o_1)=\delta_{i_1,o_1}$ (where $i_1,o_1\in\{0,1\}$) which results in the grandfather antinomy \cite{BT21}---corresponding to $p(x_1|a_1)=0$ (for all $x_1,a_1$)---for the intervention $p(x_1,o_1|a_1,i_1)=\delta_{o_1,i_1\oplus 1}\delta_{x_1,a_1}$ (where $x_1,a_1\in\{0,1\}$).}

\textit{Antinomicity.---}For nonsignaling correlations $p(\vec{x}|\vec{a})$, local causality \cite{Bell76, Wiseman14} requires that $p(\vec{x}|\vec{a})=\sum_{\vec{i}}\prod_{k=1}^Np(x_k|a_k,i_k)p(\vec{i})$, where $\vec{i}=(i_1,i_2,\dots,i_N)$ denotes a source of classical shared randomness that is distributed among the parties and $p(x_k|a_k,i_k)$ denotes the local strategy of party $S_k$ with the key feature that it is independent of the settings and outcomes of other (spacelike separated) parties. Here $p(\vec{i})$ lives in a probability simplex with the vertices of the simplex denoting deterministic assignments to $\vec{i}$.
In terms of classical processes, local causality is mathematically equivalent to requiring that the parties cannot signal to each other via the environment, \textit{i.e.},  $p(\vec{i}|\vec{o})=p(\vec{i})$ for all $\vec{o}$, so that $\vec{o}$ in Eq.~\eqref{eq:loccons1} can be marginalized and we recover correlations within the Bell polytope. Hence, $p(\vec{i})$ is a nonsignaling classical process. It can be understood as a probabilistic mixture of deterministic nonsignaling classical processes, \textit{i.e.}, $p(\vec{i})=\sum_lp(l)\delta_{\vec{i},\vec{i}_l}$, where $l$ labels deterministic assignments $\vec{i}_l$ to $\vec{i}$. This is consistent with the idea that any indeterminism in classical physical theories (like special or general relativity) can always be understood as one's lack of knowledge about an underlying physics that is deterministic. In keeping with this idea when we move to the single-round communication scenario, we propose that the most general correlations achievable in a classical physical theory without definite causal order are those that can be understood as arising from probabilistic mixtures of process functions. We refer to this notion of classicality as \textit{deterministic consistency}, or simply, \textit{nomicity}. Deterministic consistency (or nomicity) can be viewed as a conjunction of two assumptions on the realizability of a correlation via some classical quasi-process under local interventions:\footnote{As we will show further on, \textit{every} correlation admits a realization with a classical quasi-process under local interventions if no further assumptions are imposed on the realization.} firstly, that the classical quasi-process satisfies \textit{logical consistency}, \textit{i.e.}, it is a classical process, and, secondly, that it satisfies \textit{determinism}, \textit{i.e.}, it lies within the deterministic-extrema polytope \cite{BW16}.

We refer to the failure of deterministic consistency or nomicity (analogous to the failure of local causality) as \textit{antinomicity} (analogous to nonlocality), \textit{i.e.}, any correlation that fails to be nomic is \textit{antinomic}. This terminology is motivated by the fact that antinomicity entails the presence of time-travel antinomies \cite{Baumeler17, BT21} in any underlying classical explanation of the correlation.

\textit{Results.---}We can now define a hierarchy of sets of correlations as follows: i) Deterministically Consistent (nomic) correlations $\mathcal{DC}$ (achievable by convex mixtures of process functions), ii) Probabilistically Consistent correlations $\mathcal{PC}$ (achievable by classical processes), iii) Quantum Process correlations $\mathcal{QP}$ (achievable by process matrices), and iv) Quasi-consistent correlations $\mathcal{qC}$ (achievable by classical quasi-processes). Our main result establishes the following strict inclusions: 
\begin{equation}
\mathcal{DC}\subsetneq\mathcal{PC}\subsetneq\mathcal{QP}\subsetneq\mathcal{qC}.	
\end{equation}
We show in the Appendix that $\mathcal{qC}$ is the set of all multipartite correlations. The following theorem then is key to these strict inclusions.
\begin{theorem}\label{thm:det}
	Every deterministic correlation that can be realized by a process matrix can also be realized by a process function.
\end{theorem}
Theorem \ref{thm:det} follows from Theorem 4 in our companion article \cite{KO23}. We outline a proof sketch in the Appendix. Theorem \ref{thm:det} can be viewed as a generalization of the following observation that holds in Bell scenarios: every deterministic nonsignaling correlation (vertices of the Bell polytope) that can be realized by a quantum state can also be realized by a local hidden variable model. The logic of the strict inclusions is then as follows. 

1) $\mathcal{QP}\subsetneq\mathcal{qC}$: The bipartite Guess Your Neighbor's Input (GYNI) game requires each party to guess (as \textit{outcome}) the other party's input (\textit{setting}) \cite{BAF15}. A correlation that wins the GYNI game perfectly entails that for $a_1,a_2,x_1,x_2\in \{0,1\}$, the parties $S_1$ and $S_2$ should guess each other's inputs (settings) deterministically, \textit{i.e.}, $x_1=a_2$ and $x_2=a_1$. In the bipartite case, perfect GYNI correlation is unachievable by any process function since there are no causal inequality violations in the bipartite diagonal limit of the process-matrix framework \cite{OCB12}. Hence, by Theorem \ref{thm:det}, perfect GYNI correlation is impossible with process matrices, \textit{i.e.}, $\mathcal{QP}\subsetneq \mathcal{qC}$.

2) $\mathcal{PC}\subsetneq\mathcal{QP}$: This follows from the fact that in the bipartite case $\mathcal{DC}=\mathcal{PC}$ and that bipartite causal inequalities are violated by process matrices \cite{OCB12, BW16}.

3) $\mathcal{DC}\subsetneq\mathcal{PC}$: This strict inclusion follows from our construction of the tripartite Guess Your Neighbor's Input, or NOT (GYNIN) inequality and its violation, as we demonstrate below.

\textit{GYNIN game:} Three parties $S_1,S_2,S_3$ receive settings $a_1,a_2,a_3$ (respectively) and report outcomes $x_1,x_2,x_3$ (respectively) with the winning condition that either $(x_1,x_2,x_3)=(a_3,a_1,a_2)$ or $(x_1,x_2,x_3)=(\bar{a}_3,\bar{a}_1,\bar{a}_2)$. The winning probability when the settings are drawn uniformly at random is given by $p_{\rm gynin}:=\frac{1}{8}\sum_{\vec{x},\vec{a}}p(\vec{x}|\vec{a})\left(\delta_{x_1,a_3}\delta_{x_2,a_1}\delta_{x_3,a_2}+\delta_{x_1,\bar{a}_3}\delta_{x_2,\bar{a}_1}\delta_{x_3,\bar{a}_2}\right)$.

The following GYNIN inequality then serves as our witness of antinomicity:
\begin{align}\label{eq:gynin}
	p_{\rm gynin}\leq\frac{5}{8}.
\end{align}
This inequality is saturated by the deterministic AF/BW process \cite{BW16, KB23} but not by any causal strategy since the causal bound on the winning probability is $\frac{1}{2}$. Finally, this game can be won perfectly, \textit{i.e.}, with $p_{\rm gynin}=1$, for a probabilistically consistent correlation realized by the Baumeler-Feix-Wolf (BFW) process \cite{BFW14}. This establishes the strict inclusion $\mathcal{DC}\subsetneq\mathcal{PC}$. The causal bound follows quite similarly as in the case of other causal inequalities, \textit{e.g.}, GYNI inequality \cite{BAF15}. We provide a proof sketch for the classical bound in the Appendix and refer to our companion article \cite{KO23} for more details. The BFW process \cite{BFW14} can be expressed as a conditional probability distribution given by $p(\vec{i}|\vec{o}):=\frac{1}{2}\delta_{i_1,o_3}\delta_{i_2,o_1}\delta_{i_3,o_2}+\frac{1}{2}\delta_{i_1,\bar{o}_3}\delta_{i_2,\bar{o}_1}\delta_{i_3,\bar{o}_2}$, where $i_k,o_k\in\{0,1\}$ for all $k\in\{1,2,3\}$. The interventions on this process that, via Eq.~\eqref{eq:loccons1}, win the GYNIN game perfectly are given by $p(x_k,o_k|a_k,i_k)=\delta_{x_k,i_k}\delta_{o_k,a_k}$ for all $k\in\{1,2,3\}$.

\begin{figure}
	\includegraphics[scale=0.35]{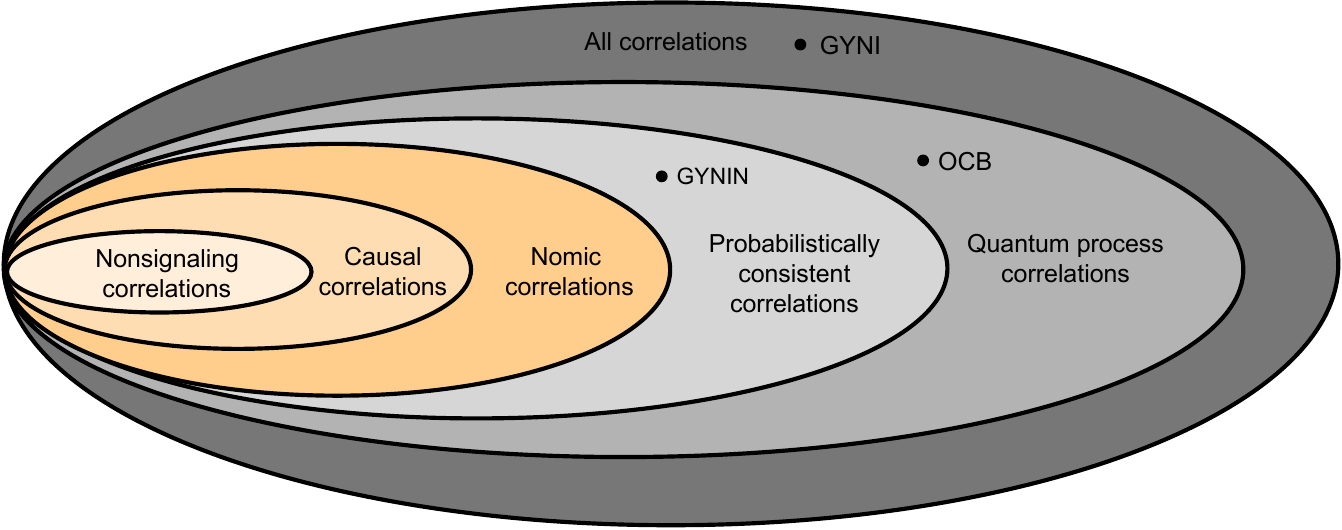}
	\caption{Inclusion relations between different sets of correlations. The point GYNIN refers to the correlation for which $p_{\rm gynin}=1$, achieved by the BFW process \cite{BFW14}. The point OCB refers to the correlation that wins the Oreshkov-Costa-Brukner game with probability $\frac{2+\sqrt{2}}{4}$ \cite{OCB12}. The point GYNI \cite{BAF15} refers to the correlation that wins the GYNI game with probability $1$. Antinomicity witnesses such as Eq.~\eqref{eq:gynin} separate nomic correlations from the rest.}\label{fig:setinclusions}
\end{figure}

\textit{Discussion.---}
Going beyond Bell and causal inequalities \cite{BCP14, OCB12}, we have introduced antinomicity as a notion of nonclassicality for correlations in single-round communication scenarios without global causal assumptions (cf.~Fig.~\ref{fig:setinclusions}). Similarly to Bell nonlocality \cite{WS15}, antinomicity requires fine-tuning \cite{WS15} underlying statistical parameters in a classical model to avoid operational time-travel antinomies. The physical implications of antinomicity for correlations in the process-matrix framework depend on the existence of a unitarily-extendible process that can exhibit antinomic correlations, \textit{e.g.}, by violating Eq.~\eqref{eq:gynin}. If such a process exists, then antinomicity of the associated correlation would certify that the correlation cannot be causally explained using \textit{classical split-node causal models} while still admitting a causal explanation via a \textit{quantum causal model} (in the sense of Ref.~\cite{BLO21}). If, on the other hand, such a process provably doesn't exist, then antinomicity would serve as a device-independent witness of non-unitarily extendible processes, \textit{i.e.}, a process matrix witnessing antinomicity would necesssarily fail to be unitarily extendible, falling outside the scope of quantum causal models \cite{BLO21}. Furthermore, in this case, if one adopts the purification postulate of  Ref.~\cite{AFN17} as a necessary condition for physical realizability, antinomicity of a correlation would certify its unphysicality by this criterion. We have largely adopted a strictly operational (or device-independent) perspective \cite{WSS20, BCP14} in this article. In our companion article \cite{KO23} we adopt a more causal perspective on antinomicity  \cite{WS15, WSS20, ABH17,BLO19,BLO21,TB23} and refer the reader to it for a deep dive in that direction.
\ack
R.K.~thanks Ämin Baumeler, Hippolyte Dourdent, Tobias Fritz, and Andreas Winter for discussions that informed his thinking while pursuing this project. This work was made possible through the support of the ID\# 62312 grant from the John Templeton Foundation, as part of the project \href{https://www.templeton.org/grant/the-quantum-information-structure-of-spacetime-qiss-second-phase}{`The Quantum Information Structure of Spacetime' (QISS)}. The opinions expressed in this work are those of the author(s) and do not necessarily reflect the views of the John Templeton Foundation. R.K.~was supported by the Chargé de Recherche fellowship of the Fonds de la Recherche Scientifique (F.R.S.-FNRS), Belgium and the Program of Concerted Research Actions (ARC) of the Université libre de Bruxelles. O.O.~is a Research Associate of the Fonds de la Recherche Scientifique (F.R.S.-FNRS), Belgium. This work was supported by the F.R.S.-FNRS under project CHEQS within the Excellence of Science (EOS) program. This work also received support from the French government under the France 2030 investment plan, as part of the Initiative d'Excellence d'Aix-Marseille Université-A*MIDEX, AMX-22-CEI-01.

\section{Appendix}

\textit{$\mathcal{qC}$ is the set of all correlations:}
To show that the set $\mathcal{qC}$ is the set of all correlations, we simply need to provide a quasi-consistent realization of any given correlation $p(\vec{x}|\vec{a})$. The following quasi-consistent realization works for any given correlation $p(\vec{x}|\vec{a})$: We define the classical quasi-process and the associated local interventions via 
\begin{align}
	&p(\vec{i}|\vec{o}):=\sum_{\vec{x}',\vec{a}'}p(\vec{x}'|\vec{a}')\delta_{\vec{i},\vec{x}'}\delta_{\vec{o},\vec{a}'},\\
	&p(x_k,o_k|a_k,i_k):=\delta_{x_k,i_k}\delta_{o_k,a_k},
\end{align}
so that 
\begin{align}
	\sum_{\vec{i},\vec{o}}\prod_{k=1}^N\delta_{x_k,i_k}\delta_{o_k,a_k}\sum_{\vec{x}',\vec{a}'}p(\vec{x}'|\vec{a}')\delta_{\vec{i},\vec{x}'}\delta_{\vec{o},\vec{a}'}=p(\vec{x}|\vec{a}).
\end{align}
Here the classical quasi-process encodes $p(\vec{x}|\vec{a})$ perfectly and the local interventions simply extract the correlations implicit in $p(\vec{i}|\vec{o})$. The correlation set $\mathcal{qC}$ is therefore subject only to the minimal constraints of non-negativity and normalization, \textit{i.e.},
\begin{align}
	p(\vec{x}|\vec{a})\geq 0\quad\forall \vec{x},\vec{a}\textrm{ and }\sum_{\vec{x}}p(\vec{x}|\vec{a})=1\quad\forall \vec{a}.
\end{align}

\textit{Proof sketch of Theorem \ref{thm:det}.---}

\begin{figure*}
	\includegraphics[scale=0.35]{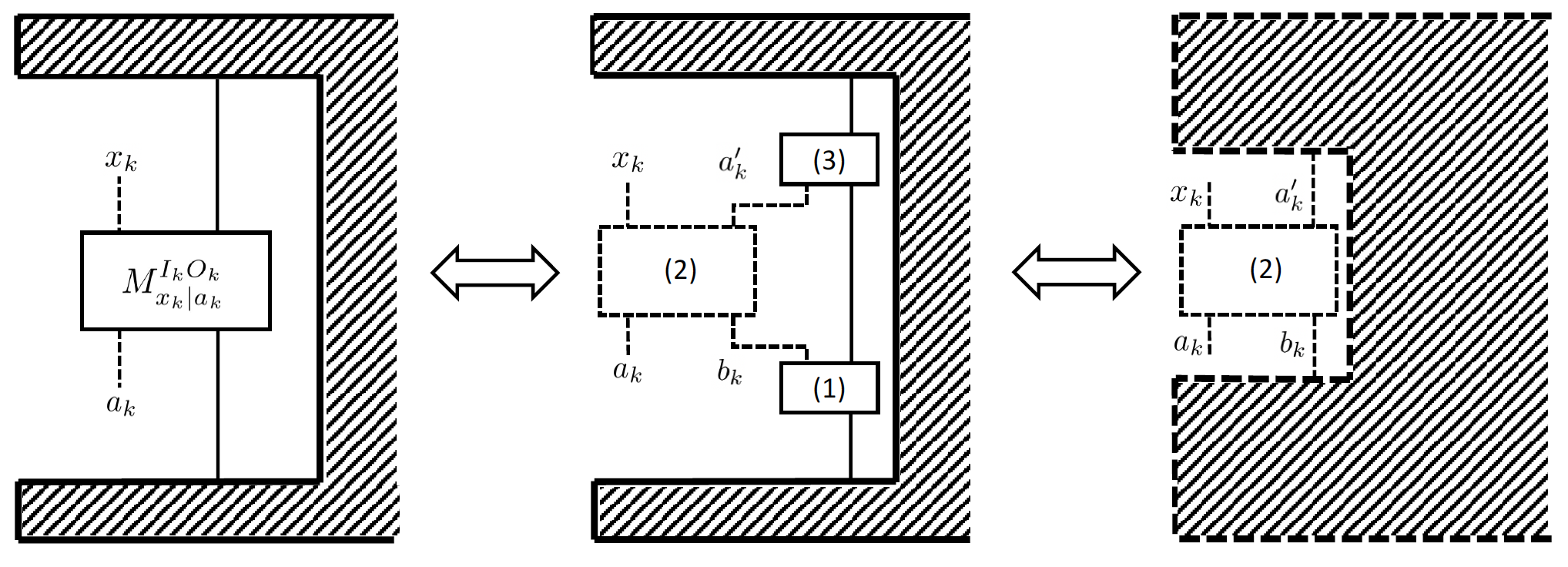}
	\caption{Transformation of the local operations of any party $S_k$ without changing the deterministic correlation $\vec{x}=f(\vec{a})$, following Lemma \ref{lem:replaceinstrument}. Solid lines indicate quantum instruments, systems, and processes, while dashed lines indicated classical instruments, systems, and processes.}\label{fig:seqofopns}
\end{figure*}

We provide a proof sketch of Theorem \ref{thm:det} below. 

As illustrated in Fig.~\ref{fig:seqofopns}, we first argue that in any process-matrix realization (using a process matrix $W$, say) of the correlation $p(\vec{x}|\vec{a})=\delta_{\vec{x},f(\vec{a})}$, we can replace the quantum instrument of any party $S_k$ ($k\in[N]$) by a sequence of three instruments without changing the correlation: 
\begin{itemize}
	\item a fixed L\"uders instrument with classical outcome $b_k$ and resulting state $\rho_{b_k}\in\mathcal{L}(\mathcal{H}^{b_k})$, 
	\item a classical instrument $p(x_k,a'_k|a_k,b_k)$, and 
	\item a fixed quantum instrument $M^{'b_kO_k}_{a'_k}$ that accepts the inputs $a'_k$ and $\rho_{b_k}$ and performs exactly as the original quantum instrument $M_{a'_k}^{I_kO_k}$, but now restricted to the subspace $\mathcal{H}^{b_k}\subseteq \mathcal{H}^{I_k}$.
\end{itemize}
This is summarized in the following lemma:
\begin{lemma}\label{lem:replaceinstrument}
	In any process-matrix realization of a deterministic correlation $p(\vec{x}|\vec{a})=\delta_{\vec{x},f(\vec{a})}$, the local quantum operation performed by any party $S_k$ (where $k\in[N]$) can be replaced by a sequence of operations as described in Fig.~\ref{fig:seqofopns} without changing the deterministic correlation, where
	\begin{itemize}
		\item 	the instrument (1) is a fixed Lüders instrument that projects the incoming state onto one of a set of mutually orthogonal subspaces $\mathcal{H}^{b_k}$, where $b_k\in\{1_k,\dots,N_k\}$, yielding classical outcome $b_k$, and outputting the projected state  $\rho_{b_k}\in\mathcal{L}(\mathcal{H}^{b_k})\subseteq\mathcal{L}(\mathcal{H}^{I_k})$;
		
		\item the instrument (2) is a classical instrument dependent on the local setting $a_k$ that takes as an input the outcome $b_k$ of instrument (1), sends out the output $a'_k = a_k$, and produces the outcome $x_k = f'_k(a_k, b_k)$;
		
		\item the instrument (3) is a fixed instrument $M^{'b_kO_k}_{a'_k}$ that takes as input $a'_k$ and $\rho_{b_k}$ and performs exactly what the original quantum instrument $\{M^{I_kO_k}_{x_k|a_k}\}_{x_k\in X_k}$ would perform from $\mathcal{H}^{I_k}$ to $\mathcal{H}^{O_k}$ depending on $a_k$, but now from $\mathcal{H}^{b_k}$  to $\mathcal{H}^{O_k}$ conditionally on the value of $a’_k$, with the outcome of that operation traced out.		
	\end{itemize}
\end{lemma}
We refer the reader to our companion article \cite{KO23} for a complete proof of Lemma \ref{lem:replaceinstrument} and simply outline below the logic of the remaining steps towards proving Theorem \ref{thm:det}.

\begin{enumerate}
	\item Using Lemma \ref{lem:replaceinstrument}, we replace the local intervention of each party by those in Fig.~\ref{fig:seqofopns}. This yields a realization of $p(\vec{x}|\vec{a})$ using the same process matrix $W$ as before but now with local  interventions of the type illustrated in Fig.~\ref{fig:seqofopns}.
	
	\item To show that $p(\vec{x}|
	\vec{a})$ can also be obtained by applying local interventions on a classical process, we argue that we can absorb instruments (1) and (3) in the surrounding process, thereby redefining the local lab---as one that implements only the classical instrument (2)---as well as the surrounding process. 
	
	\item The reason the above procedure results in a valid process is the following: in the original lab one could replace a particular classical instrument (2) with any other classical instrument and still obtain a valid correlation since that would still define an overall instrument (sequentially composed of (1), (2), and (3)) in the original lab. Since the new lab has (classical) input $b_k$ and (classical) output $a'_k$, the new process we obtain through this procedure is a classical process. 
	
	\item We then argue---based on our construction of instrument (1) while proving Lemma \ref{lem:replaceinstrument}---that the classical process is, in fact, a process function, \textit{i.e.}, any string $\vec{a'}$ sent out by the parties to the process is mapped to a unique string $\vec{b}$ received by the parties.
\end{enumerate}
This completes our proof sketch for Theorem \ref{thm:det} and we refer the reader to our companion article \cite{KO23} for more details.

\textit{Proof sketch of the GYNIN bound in Eq.~\eqref{eq:gynin}.---}
To obtain the classical upper bound, we first argue that we can, without loss of generality, consider process functions with binary inputs and outputs while maximizing the value of $p_{\rm gynin}$ over process functions. This simplifies the problem to simply maximizing $p_{\rm gynin}$ over variants of the AF/BW process following the characterization of Baumeler and Wolf \cite{BW16}, yielding the upper bound of $\frac{5}{8}$.

\bibliography{masterbibfilev2.bib}

\end{document}